\documentclass[pdflatex]{article}
\usepackage{graphicx}
\usepackage{multirow}%
\usepackage{amsmath,amssymb,amsfonts}%
\usepackage{amsthm}%
\usepackage{mathrsfs}%
\usepackage[title]{appendix}%
\usepackage{xcolor}%
\usepackage{textcomp}%
\usepackage{manyfoot}%
\usepackage{booktabs}%
\usepackage{algorithm}%
\usepackage{algorithmicx}%
\usepackage{algpseudocode}%
\usepackage{listings}%
\usepackage{authblk}

\begin{document}

\title{Software development projects as a way for multidisciplinary soft and future skills education.}
% \runtitle{Software development projects as a way for multidisciplinary soft and future skills education.}
%Teaching soft and future skills to computer science students through international interdisciplinary projects

\author[1]{Krzysztof Podlaski}% \email{krzysztof.podlaski@uni.lodz.pl}}
\author[1]{Michał Beczkowski}% \email{michal.beczkowski@uni.lodz.pl}}
\affil[1]{University of Lodz, Poland}

\author[2]{Katharina Simbeck}% \email{katharina.simbeck@htw-berlin.de}}\equalcont{These authors contributed equally to this work.}
\author[2]{Katrin Dziergwa}% \email{katrin.dziergwa@htw-berlin.de}}\equalcont{These authors contributed equally to this work.}
\affil[2]{HTW University of Applied Science, Germany}
\author[3]{Derek O'Reilly}% \email{derek.oreilly@dkit.ie}}\equalcont{These authors contributed equally to this work.}
\author[3]{Shane Dowdall}% \email{shane.dowdall@dkit.ie}}\equalcont{These authors contributed equally to this work.}
\affil[3]{Dundalk Institute of Technology, Ireland}
\author[4]{Joao Monteiro}% \email{jmonteiro@ispgaya.pt}}\equalcont{These authors contributed equally to this work.}
\author[4]{ Catarina Oliveira Lucas}% \email{clucas@ispgaya.pt}}\equalcont{These authors contributed equally to this work.}
\affil[4]{Instituto Superior Politécnico Gaya, Portugal}
\author[5]{Johanna Hautamaki}% \email{johanna.hautamaki@centria.fi}}\equalcont{These authors contributed equally to this work.}
\author[5]{Heikki Ahonen}% \email{heikki.ahonen@centria.fi}}\equalcont{These authors contributed equally to this work.}
\affil[5]{Centria University of Applied Science, Finland}
\author[6]{Hiram Bollaert}% \email{hiram.bollaert@ap.be}}\equalcont{These authors contributed equally to this work.}
\author[6]{Philippe Possemiers}% \email{philippe.possemiers@ap.be}}\equalcont{These authors contributed equally to this work.}
\affil[6]{AP University of Applied Science and Art, Belgium}
\author[7]{Zofia Stawska}% \email{zofia.stawska@p.lodz.pl}}\equalcont{These authors contributed equally to this work.}
\affil[7]{Lodz University of Technology, Poland}
\date{}
\maketitle

\abstract{
  %{\bf Background:}
  Soft and future skills are in high demand in the modern job market. These skills are required for both technical and non-technical people. It is difficult to teach these competencies in a classical academic environment.

  %{\bf Objectives:}
  The paper presents a possible approach to teaching in soft and future skills in a short, intensive joint project. In our case, it is a project within the Erasmus+ framework, but it can be organized in many different frameworks.

  %{\bf Methods:}
  In the project we use problem based learning, active learning and group-work teaching methodologies. Moreover, the approach put high emphasizes diversity. We arrange a set of multidisciplinary students in groups. Each group is working on software development tasks. This type of projects demand diversity, and only a part of the team needs technical skills. In our case less than half of participants had computer science background. Additionally, software development projects are usually interesting for non-technical students.

  %{\bf Results:}
  The multicultural, multidisciplinary and international aspects are very important in a modern global working environment. On the other hand, short time of the project and its intensity allow to simulate stressful situations in a real word tasks. The effects of the project on the required competencies are measured using the KYSS method.

  %{\bf Conclusions:}
  The results prove that the  presented method increased participants soft skills in communication, cooperation,  digital skills and self reflection.

{\bf keywords: } soft-skills, future-skills, intensive, multidisciplinary, international, intercultural, active-learing

}

\section{Introduction}
Soft and future skills are vital for graduates \cite{oecd2018, ehlers2019}. However, many students underestimate their importance and believe that domain competencies are sufficient \cite{Succi2019}. Students who recognize the importance of soft skills achieve higher salaries, while graduates with lower wages focus primarily on hard skills \cite{Lamberti2021}. Additionally, the authors of \cite{Gnecco2023} predict that after COVID-19, the changes in working conditions will further increase the demand for soft skills of employees. Higher education institutions (HEIs) need not only to provide students with appropriate domain competencies, but also help them develop inter-domain and interpersonal skills. Therefore, HEIs have to modify teaching methodologies accordingly. Many approaches have been proposed to address the mentioned demands. Most of them are based on active learning, problem-solving, and group-work methods \cite{VandenBeemt2020, Chen2020, EstradaMolina2022, Whewell2022}.

Active learning and problem-based methods are easily applicable in an educational institution, but these methods alone are not enough to increase cooperation and communication skills like empathy. %find citations about limitations to teach those in HEIs
Moreover, in real-life working environment, the IT specialists often have to cooperate with non-technical persons \cite{Frezza2019}.  Unfortunately, incorporation of multidisciplinary and intercultural aspects into teaching in class environments is not straightforward.
In this paper we present an approach to teaching soft and future skills in an international cooperation project between HEIs and analyse its effect on students' self-assessed competencies with regards to communication, cooperation, flexibility, digital skills, creativity, critical thinking, willingness to learn and self-reflection. In the project, during a 10-day intensive course, students worked in interdisciplinary teams with other students from different (academic) cultures, different languages and different backgrounds. The result on increased soft and future skills in participating students is measured using the KYSS questionnaire \cite{Chaoui2022, DeBruyne2023}. The development of KYSS (see Section \ref{sec:kyss}) was focused on correspondence between survey questions, language and its understanding by participants as well as on measurement of soft and social skills. %add reference

The paper is structured as follows. In the Section \ref{sec:related_works} we discuss the problem of soft- and future-skills in modern education. Next section is dedicated to description of the project. Later we introduce KYSS survey. Sections \ref{sec:assesment}  and \ref{sec:discussion} focus on results and interpretation of KYSS surveys conducted during the project. At the end of the paper we present Conclusions.

\section{Related Works} \label{sec:related_works}
\subsection{Soft and future skills}
%soft skills
The importance of soft skills has long been established \cite{kechagias2011teaching, schulz2008importance}. While hard skills usually refer to domain specific competencies, such as databases, programming or operating systems, soft skills are complementary skills that are important for professional success and operating in a team in working environment  \cite{schulz2008importance, cimatti2016definition}. While there is no consensual definition of soft skills nor a finite list of single skills, soft skills usually include skills related to interpersonal skills (communication, collaboration, empathy), reflection (critical thinking, problem solving) and self-development (learning, self-reflection) \cite{schulz2008importance, cimatti2016definition}. Soft skills are hard to measure, they are best learned when they are integrated with hard skills \cite{cimatti2016definition}.
%future skills
The concept of future skills or 21st century skills extends the concept of soft skills towards skills that help to cope with an ever changing professional environment. Future skills overlap with soft skills, but prioritize information management, critical thinking, creativity, problem solving, collaboration, communication, self-direction, lifelong learning, ethical awareness, cultural awareness and flexibility \cite{van2017relation, ananiadou200921st}. Like soft skills, future skills are difficult to measure, too \cite{ananiadou200921st}. As additional requirement for future workforce is entrepreneurial readiness.  Entrepreneurship and its connection with computational thinking is important a much computer science as well as in social study disciplines \cite{Kang2020}.

The soft skills most frequently addressed in curricula are communication, teamwork, ethics and presentation skills \cite{groeneveld2020soft}. However, skills such as creativity and empathy are lacking in the curricula, even though they are required by industry \cite{groeneveld2020soft}. Important future skills such as critical thinking and self-reflection were only found in $16\%$ and $13\%$ of the analysed curricula respectively \cite{groeneveld2020soft}.

% specific skills: communication, cooperation, flexibility, digital skills, creativity, critical thinking, willingness to learn, self-reflection
%communcation skills
Probably the most frequently cited soft skill is the ability to communicate. This includes the ability to communicate in English as the lingua franca (including the ability to communicate with non-native speakers), to communicate across disciplines, hierarchies, cultures and genders, in written and oral form, and to listen and to visualize \cite{riemer2007communication, balaji2009comparative}. Communications skills are often taught using presentations, peer review, role play or team tasks \cite{riemer2007communication}. Communication skills are especially important in situations such as teamwork, negotiations, job interviews and mentoring \cite{hargie1997handbook}.

%cooperation
The ability to jointly solve a task with others is referred to as cooperation skills. Problem based learning \cite{panlumlers2015design, trisdiono2019development} and portfolio projects \cite{banhegyi2023improving} have been documented to improve cooperation skills.

%flexibility
Flexibility is the ability to adapt to changing circumstances. It is especially relevant for project management, for example in software development \cite{sukhoo2005accommodating}. Flexibility includes the ability to change, to learn, to accept and to adjust \cite{robles2012executive, balaji2009comparative}.

%digital skills
Digital skills refer to the proficient but ciritical use of digital information, media and tools \cite{sicilia2018digital} as well as the attitude towards those \cite{van2017relation}.
%creativity
Creativity is the ability to generate new ideas or recombine existing ideas into new concepts \cite{heye2006creativity}. Learning creativity is associated with risk taking, diversity of inputs, generation of ideas and evaluating/prioritizing them later \cite{heye2006creativity}.

%critical thinking
Critical thinking as a skill has been studied extensively since the 1990s \cite{behar2011teaching}. Critical thinking is related to problem solving and requires domain knowledge and skills as a base for intellectual and cognitive analysis, interpretation, argumentation and judgement \cite{behar2011teaching, alsaleh2020teaching}. Critical thinking appears to be more difficult to teach and requires longer interventions in comparison with other skills \cite{behar2011teaching}. Because of the link to domain knowledge, critical thinking skills improve with time spent studying and books read \cite{terenzini1995influences}.

%willingness to learn
Willingness to learn is the skill that includes the notions of being responsible for one's own learning, self-management of learning and self-expertise of learning \cite{simons2008our}. Willingness to learn requires alertness, openness and reflection, as learning does not always take place automatically, through experience \cite{van2006exploring}.

%self-reflection
Self-reflection involves the purposeful mental processing of one's own learning process and outcomes. It is a pre-requisite for problem-based learning \cite{lew2011self}.
Self-reflection is especially important when learners face anxiety in complex tasks in team projects \cite{schulz2023social}. Self-reflection is often taught through reflective journal writing \cite{black2000learning,lew2011self}.

In education, soft skills can be taught using didactic settings that require teamwork, reflection and diversity \cite{hazzan2013teaching}.

\subsection{Active and problem based learning}
Active learning and problem based methods are commonly used in education, especially in education. Many classes end with real-life small projects and probably all higher education institutions use project based classes in their curricula. In ACM and IEEE Join Task Force documents we can find explicit suggestions to enforce the growth of problem based and active learning methods at least with use of  teamwork and group projects in study programs \cite{Force2013, Force2020}. In many papers, we can see how people use these approaches in teachings (see implementation review papers \cite{VandenBeemt2020, Chen2020, EstradaMolina2022, Angelaki2023} ).
Problem based learning was also successfully applied in international contexts before \cite{badets2017cross}.

It is, however, challenging to design collaborative learning settings based on learning objectives, learner characteristics and contextual factors and to align the assessment \cite{schulz2023exploring}.
Most of the implementations are in-class solutions. Therefore, they use homogeneous students, the students represent the same domain or institution. We are interested in approaches that incorporate multidisciplinary and intercultural groups with active and problem based learning. In the literature not many methods fulfill these requirements \cite{Ndiaye2023}.

In previous years our group have participated in a few joint projects \cite{OReilly2015, Milczarski2018, Monteiro2019} that lead to the development of MIMI methodology \cite{Dowdall2021}. In this project we decided to continue work with this approach as it fits our needs. MIMI acronym comes from Multinational, Intercultural, Multidisciplinary and Intensive. It is developed especially for short term  intensive projects for diverse set of participants. It provide a schedule for such an event, with detailed plan how to organize the work of teams. The organization of the event is suited to enhance selected sills. The connection of activities with expected pedagogical outcomes is sketched in the paper \cite{Dowdall2021}. The method was recognized and recommended by the authors of \cite{EstradaMolina2022}. Up to now no objective measurements of the effects were conducted. The measurements of effectiveness of small didactic experiments is biased by the size of the group. On the other hand, the idea of organizing events outside of general curricula leads to small number of participants, as such events can't be organized in bigger scale.

%Therefore, we decided to follow recommendation form \cite{EstradaMolina2022} and used the . The MIMI approach promises to fit our needs. However, objective measurements of its effect were not published.

\section{Intensive project description}\label{sec:intensive_projects}
Within an Erasmus+ cooperation partnership, six European HEIs developed and implemented an intensive 10-day course to teach soft and future skills. The methodology implemented important elements of the MIMI methodology\cite{Dowdall2021} and developed it further.  The participating 60 students and 10 lecturers met at one of the HEIs and represented different fields of study: less than half have computer science background, others were pursuing degrees in management, tourism, chemistry and production engineering. Students were assigned to teams of 6 persons with members from the six participating HEIs and at least three disciplines. The task of the groups was to create a prototype of an application or service connected with the event theme: {\em Digital Entrepreneurship and the Climate}. As a result, groups had to prepare a working prototype focused on a real life local needs, create a business potential assessment, and present the idea to internal and external audience. The organizers assigned one staff member for each group to be their mentor. The role of the mentor was to support the team and take the role of an advisor. The teams should be self-driven, and all the decisions were to be made by the student members.
The intensive course was split into three stages; each stage ended with a group presentation. The first two days are devoted to team building and brainstorming. On the second day, the teams presented ideas as a pitch speech. The second part was dedicated to the development of the idea. The teams worked on prototypes and development of the application's content and on business elements, such as stakeholders, user personas, cost assessments, Business Model Canvas, and SWOT analysis. On the fifth day of the event the groups presented their proposition again. During the last stage, the teams polished their business ideas and prototypes. On the ninth day of the project, more formal presentations took place with invited external partners who also provided independent feedback to the teams. In all presentations during the project, every member of the team had to take an active role.
Soft and future skills were self-assessed by students using the KYSS questionnaire \cite{Chaoui2022}.

Students filled out the questionnaire on the first and the last day of the intensive course.
An important difference to the MIMI methodology \cite{Dowdall2021} was that we did not conduct  lectures and workshops during the intensive course. However, short motivating, interactive sessions, similar to TED talks aligned the students' knowledge on project goals, teamwork, software and service design as well as business potential assessment. These talks allowed participating students to get acquainted with staff members who could help them later with specific problems in the course of the project.

The course was not set up as a contest, there was no winning team. It was explicitly suggested that teams help each other if possible. At the beginning of each day, the assigned mentor met with the team to conduct an assessment of tasks done, plan for the day and next days, and discuss the idea developed by the team. Often, mentors met with their teams several times per day. The frequency and length of the meetings depended on the stage of the project and the needs of the team. In the first days, the mentor helped moderate brainstorming or improve focus on idea development. Later the mentor coached the team to finish the  tasks that allow the team to achieve its goals.

\section{KYSS survey}\label{sec:kyss}

The design and assessment of didactic methodologies requires qualitative and quantitative evaluation of the results \cite{McKenney2018}. In our project we need a measurement tool that can help to measure the impact on participant's soft and future skills. The measurement of soft and future skills is not easy and obvious. The KYSS survey  \cite{Chaoui2022} fits our needs. Therefore we have decided to use this approach. The method allows to measure the level of soft and future skills in selected categories.

KYSS comes from {\it Kickstart Your Soft Skills} and was developed within the European Social Fund project under that name. The approach divides soft skills into four domains: interaction, problem-solving, information processing, and personal. For each domain some categories, like communication, cooperation, critical thinking, etc., were defined. The creators of KYSS survey developed a self-report questionnaire to estimate soft  and future skills. It can be used as self-assessment tools for everyone.  The questionnaire results in a score that is recorded and described in an individual feedback report. In addition to the scores, this feedback report also contains score-based feedback for each of the recorded skills. In the process they used standardization and validation procedures  to estimate importance of the answers and correspondence to measured skills \cite{Chaoui2022, DeBruyne2023}. The KYSS survey was prepared in dutch. At first the creators developed a vocabulary that correspond with selected soft skills. On that base they created a set of questions, and measured if a survey participant understand the question in a proper way. The work with local government institutions for unemployed people as well as VDAB one of the biggest recruitment company in Belgium allowed to build adequate and verified questionnaire. All questions are connected to selected soft-skills and the correlation was assessed in real-life environment. In our case we use translation of original dutch questions into English.

KYSS allows respondents to self-assess those skill categories using a survey questionnaire. In this project, we have decided to measure seven selected categories: communication, cooperation, digital skills, creativity, critical thinking, willingness to learn, and self-reflection.

The KYSS survey can be performed as an online survey. We asked all students to do the survey twice, at the beginning and at the end of the 10-day intensive course. We applied statistical tests to to the data to determine if there was a significant statistical difference between pre- and post-tests. The difference would suggest that there is an effect on participants' skills. Moreover, we were interested to see if the post-test results were better than the pre-test ones. Therefore, we used a one-side statistical test in order to evaluate if improvement in selected skills can be observed.

\section{Assessment of project effects}\label{sec:assesment}

During the project, we assessed the project's effects in two ways. We asked the participants to fill out a simple questionnaire to map students' reception of the project. Additionally, we used the KYSS method to measure the impact on students' soft and future skills.

In the basic questionnaire, the participants generally expressed positive reactions to the event (Tab. \ref{tab:basic_questionaire}).

\begin{table}[ht!]
    \centering
    \begin{tabular}{p{.6\textwidth}|c}
         Question&  Yes [$\%$] \\\hline
         Do you feel that this project improved your communication skills?& $96\%$\\\hline
         Do you feel that this project improved your critical thinking?& $89\%$\\\hline
         Do you feel that this project improved your creativity? & $83\%$ \\\hline
         Do you feel that the project improved your entrepreneurial skills? & $76\%$\\\hline
         Do you feel this project promotes excellence in learning, teaching and skills development?&$81\%$\\\hline
         Do you feel that this project promotes internationalization?& $96\%$\\
    \end{tabular}
    \caption{Students answers to basic questionnaire}
    \label{tab:basic_questionaire}
\end{table}

As we can see, the students evaluated the learning experience and learning gains very positively. The KYSS survey provides a deeper and more reliable understanding of the project's effects on participants' soft skills.

During the event, we measured students' soft and future skills with the use of KYSS surveys. One survey was filled out on the first day of the project, the other on the last one. Both answers were connected, and we were able to pair up the pre- and post-event answers. Students answered the questions on a five-level Likert scale: {\it 'Strongly agree', 'Agree', 'Neither agree nor disagree', 'Disagree', 'Strongly disagree'}. We assign a value of 2,1,0,-1, and -2 for each answer. This allows us to apply statistical tests and measure hypothesis if the project improves students' skills.

We could have analyzed the effect on the basis of individual questions, but we decided to work on categories, as the questions individually are less informative than when combined together. In order to assess the effect on a given category, we have summed up all the answers of a given student. We can say that the students obtained some points for answers in each of the categories for pre-and post-tests.  For each category, we have the Null Hypothesis $H_0$ that both sets of results for a given category have the same statistics. Additionally, we believe that the results of post-event surveys are better than in pre-test. We test the hypothesis $H_1$ that the results for a given category of post-event tests are statistically better than those from pre-event tests. In all categories the results of post-tests have higher mean and median than in pre-test, but it is not enough to validate the increase of skills. Therefore, we use statistical test to verify our hypothesis. One-sided rank statistical tests are much more appropriate for hypothesis testing in similar scientific problems \cite{Blair1980, conover1999}. In this research, we use the Wilcoxon one-side test \cite{Wilcoxon1945, Cureton1967},  implemented in python scipy library \cite{Scipy2020S}. For all tests, we used significance level $\alpha$ equal to $0.05$, which means if the p-value obtained for a given statistical test is lower than the significance level, we can reject the null hypothesis in favor of hypothesis $H_1$. On the other hand, if the p-value is higher than the significance level, we have no reason to reject the null hypothesis $H_0$

We present results for all categories independently.

\subsection{Category: communication}
\begin{figure}[ht!]
    \centering
    \includegraphics[width=\textwidth]{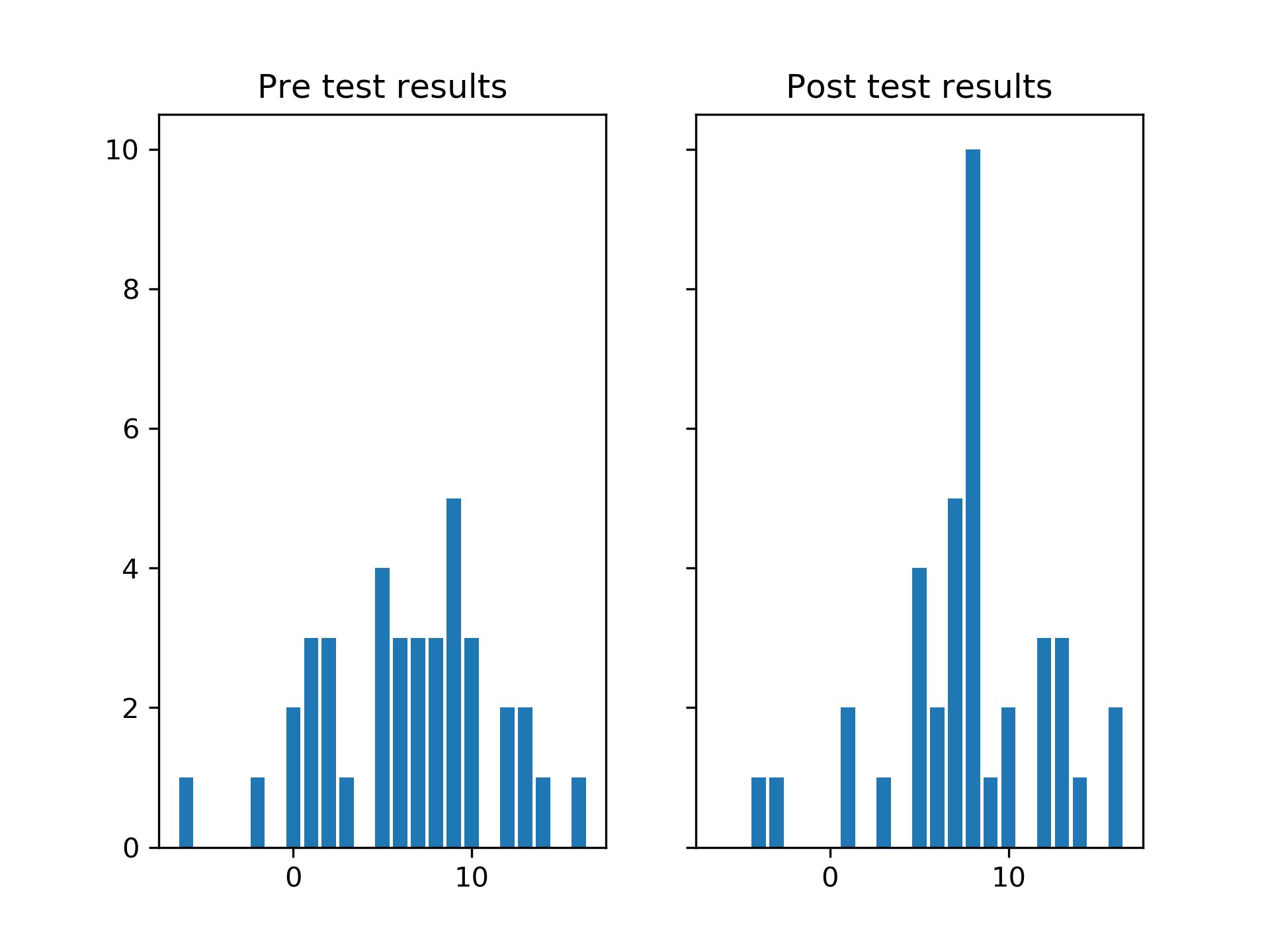}
    \label{fig:category_comm}
    \caption{Results for category communication.}.
%    }\end{minipage}
\end{figure}

This category contains eight questions. All answers for a student were summed up and paired. The results of pre-and post-survey results and the difference post-result minus pre-result are presented in Fig.~\ref{fig:category_comm}. Parameters of distribution of the results are in Table ~\ref{tab:category_comm}.

\begin{table}[ht!]
\centering
\begin{tabular}[c]{l|c|c}%
Parameter&pre&post\\%
\hline%
mean&6.368&7.763\\%
\hline%
median&7.000&8.000\\%
\hline%
$\sigma$ (st. deviation)&4.737&4.386\\%
\hline\hline
\multicolumn{3}{c}{Wilcoxon test results}\\
\hline
p-value &\multicolumn{2}{c}{0.020}\\%
\hline%
\end{tabular}%
\caption{Results for the category: communication.}
\label{tab:category_comm}
\end{table}

Using the Wilcoxon one-side test, we obtained a p-value equal to $0.020$. As this is lower than $0.05$, we can reject $H_0$ hypothesis. Therefore we accept the $H_1$ hypothesis that the results of the post-event test are statistically better than those obtained by a student in the pre-event test.

\subsection{Category: cooperation}
\begin{figure}[ht!]
    %\hfill
    %\begin{minipage}{.22\textwidth}
    %\subfloat[Cooperation]{
    \includegraphics[width=\textwidth]{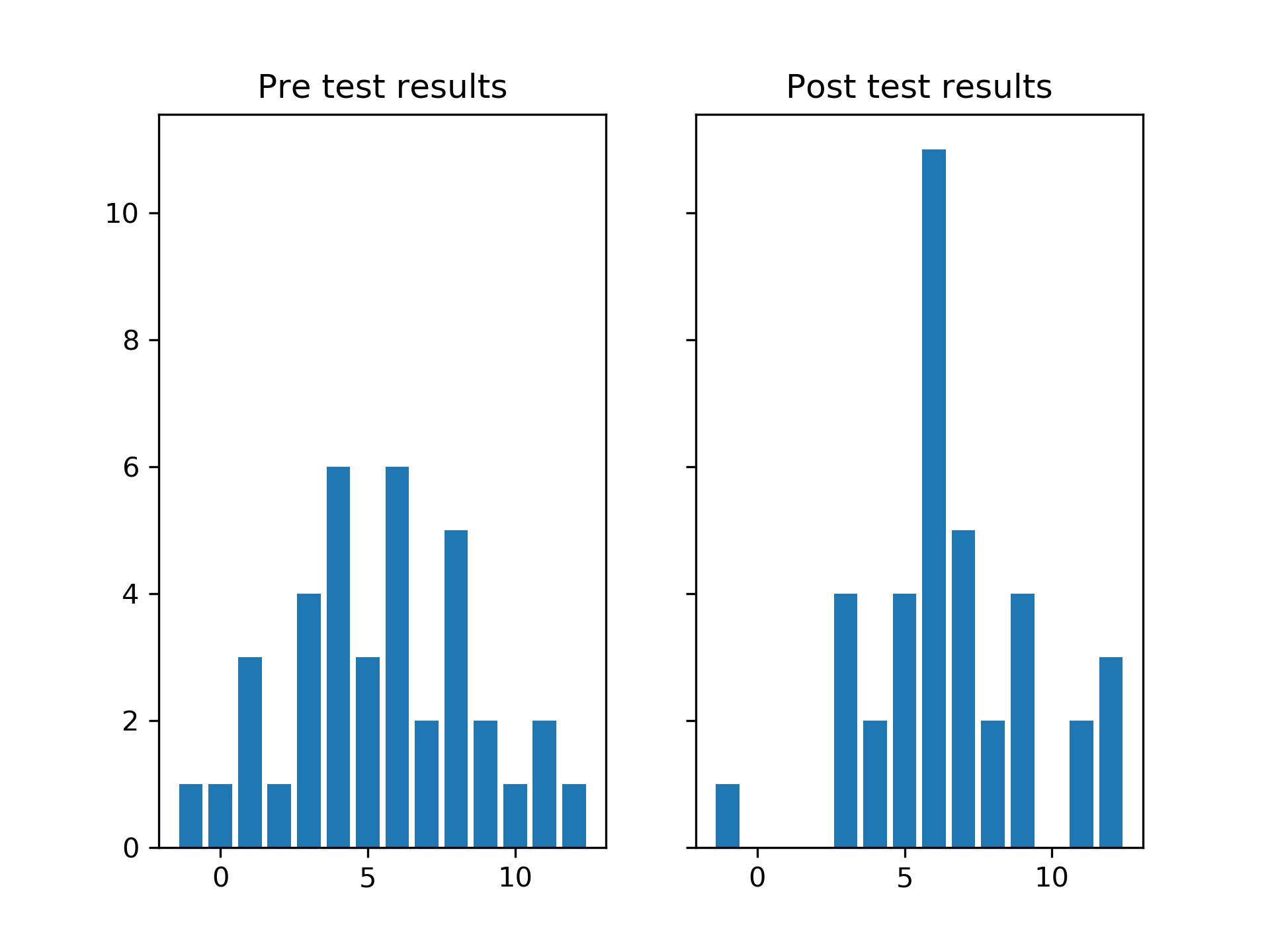}
    \label{fig:category_coop}
%    }\end{minipage}
    \caption{Results for category: cooperation.}
\end{figure}

The category cooperation contains six questions; we have proceeded in the same way as in the previous category (see Fig.~\ref{fig:category_coop} and Tab.~\ref{tab:category_coop}).
% \begin{figure}[ht!]
%     \centering
%     \includegraphics[width=.4\textwidth]{cat_coop.jpg}
%     \caption{Results for the category: cooperation.}
%     \label{fig:category_coop}
% \end{figure}

\begin{table}[ht!]
\centering
\begin{tabular}[c]{l|c|c}%
Parameter&pre&post\\%
\hline%
mean&5.447&6.579\\%
\hline%
median&5.500&6.000\\%
\hline%
$\sigma$ (st. deviation)&3.118&2.769\\%
\hline\hline
\multicolumn{3}{c}{Wilcoxon test results}\\
\hline
p-value &\multicolumn{2}{c}{0.012}\\%
\hline%
\end{tabular}%
\caption{Results for the category: cooperation.}
\label{tab:category_coop}
\end{table}

The Wilcoxon test returned a p-value equal to $0.012$, and as before, we can accept the $H_1$ hypothesis for this category.

\subsection{Category: flexibility}
\begin{figure}[ht!]
    \centering
%\begin{minipage}{.22\textwidth}
    %\subfloat[Flexibility]{
    \includegraphics[width=\textwidth]{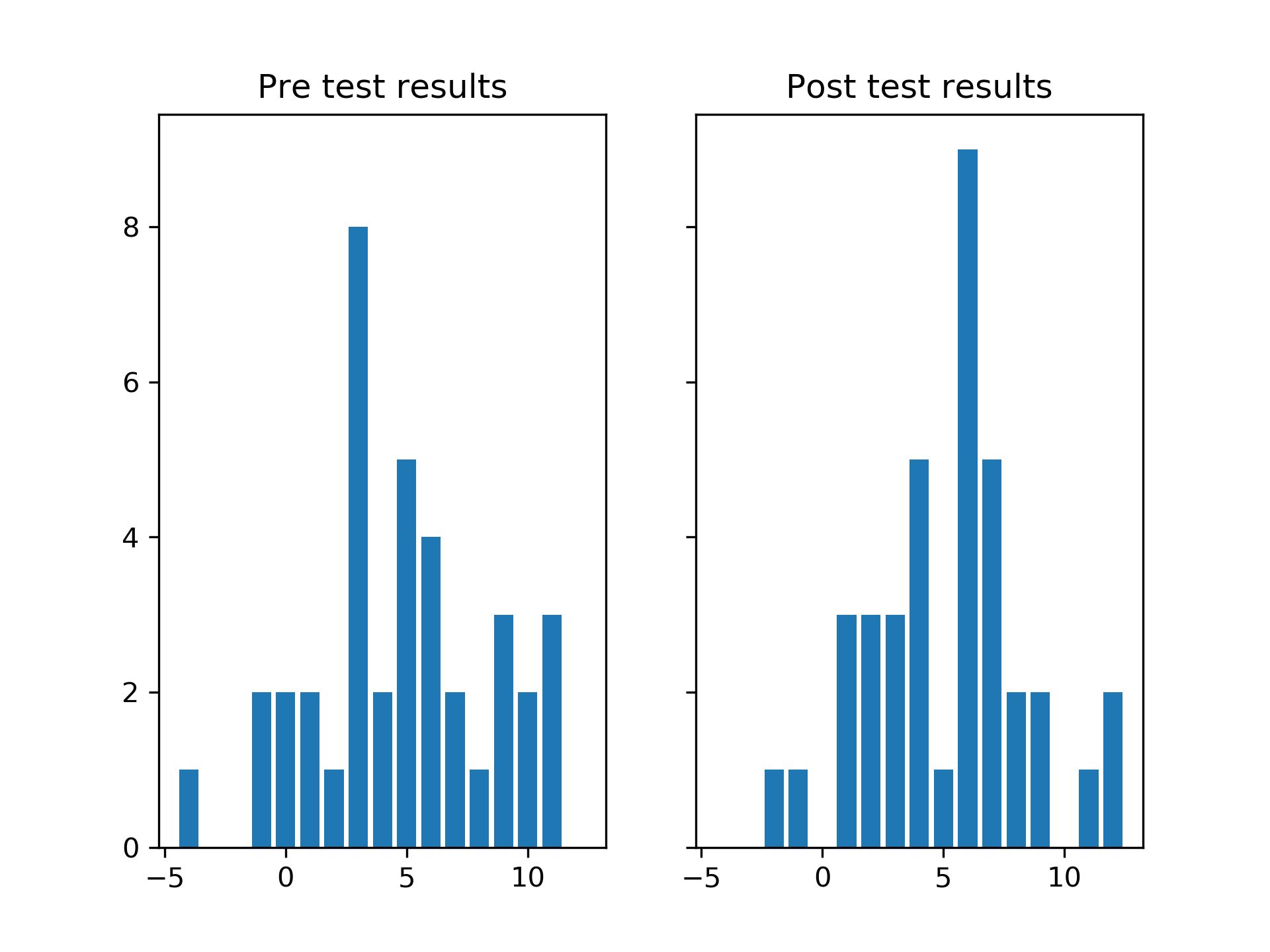}
    \label{fig:category_flex}
    \caption{Results for categories: flexibility.}
%    }\end{minipage}
    %\hfill
\end{figure}

The category flexibility contains six questions (results in Fig.~\ref{fig:category_flex}, Tab.~\ref{tab:category_flex}).

% \begin{figure}[ht!]
%     \centering
%     \includegraphics[width=.4\textwidth]{cat_flex.jpg}
%     \caption{Results for the category: flexibility.}
%     \label{fig:category_flex}
% \end{figure}
The Wilcoxon test returned a p-value equal to $0.117$, so we can't accept the $H_1$ hypothesis for this category. The conclusion is that the distribution of pre and post-results are statistically similar and can represent the same background statistics.

\begin{table}[ht!]
\centering
\begin{tabular}[c]{l|c|c}%
Parameter&pre&post\\%
\hline%
mean&4.763&5.211\\%
\hline%
median&5.000&6.000\\%
\hline%
$\sigma$ (st. deviation)&3.638&3.197\\%
\hline\hline
\multicolumn{3}{c}{Wilcoxon test results}\\
\hline
p-value &\multicolumn{2}{c}{0.117}\\%
\hline%
\end{tabular}%
\caption{Results for the category: flexibility.}
\label{tab:category_flex}
\end{table}

\subsection{Category: digital skills}
\begin{figure}[ht!]
    \centering
%    \begin{minipage}{.22\textwidth}
%    \subfloat[Digital skills]{
\includegraphics[width=\textwidth]{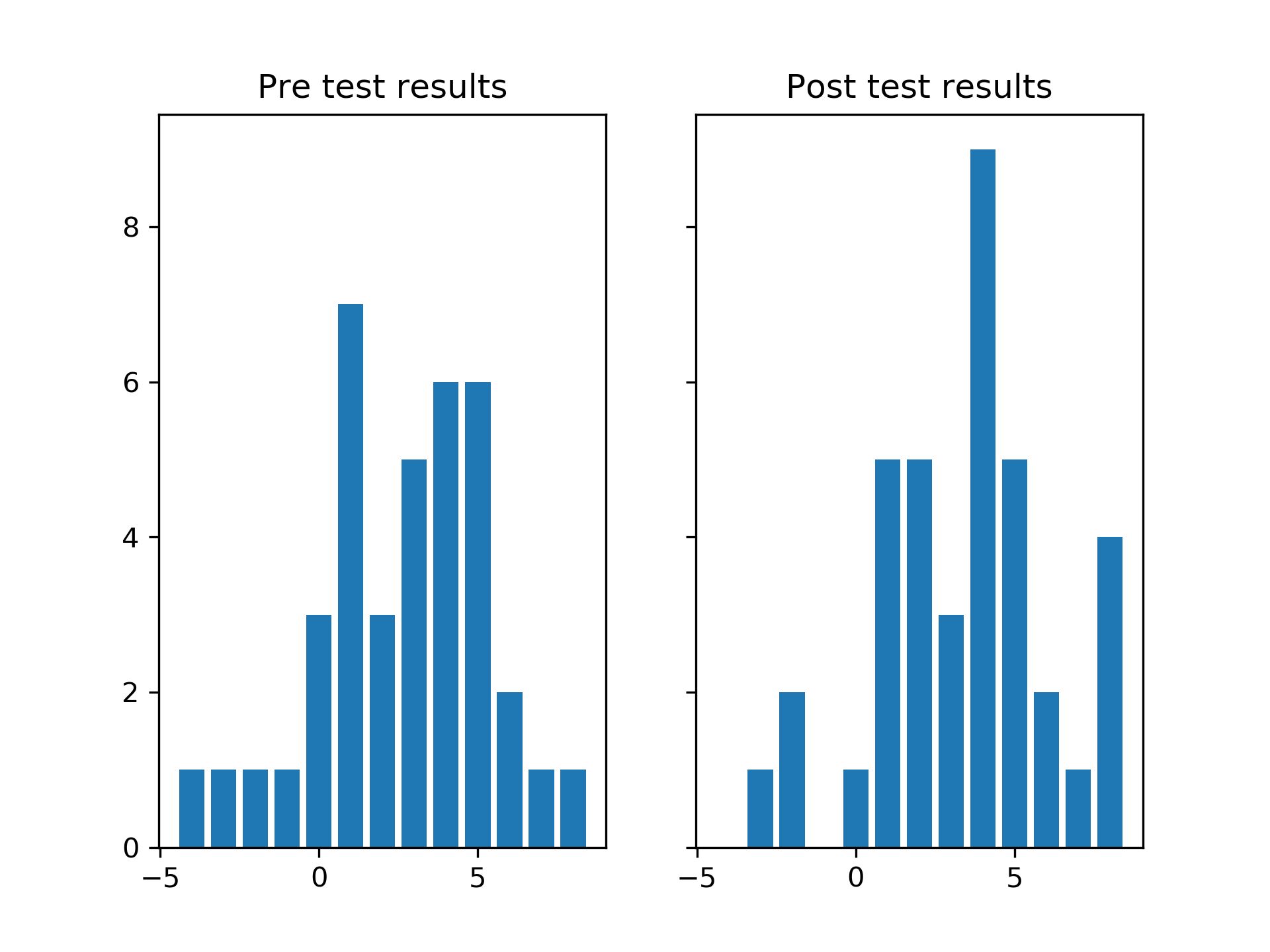}
    \label{fig:category_dig_sk}
%    }\end{minipage}
    \caption{Results for categories: flexibility and digital skills.}
\end{figure}

The category digital skills contains four questions (results in Fig.~\ref{fig:category_dig_sk} and Tab.~\ref{tab:category_dig_sk}).
% \begin{figure}[ht!]
%     \centering
%     \includegraphics[width=.4\textwidth]{cat_dig_sk.jpg}
%     \caption{Results for the category: digital skills.}
%     \label{fig:category_dig_sk}
% \end{figure}
The Wilcoxon test returned a p-value equal to $0.013$, and we can accept the $H_1$ hypothesis for this category.

\begin{table}[ht!]
\centering
\begin{tabular}[c]{l|c|c}%
Parameter&pre&post\\%
\hline%
mean&2.605&3.395\\%
\hline%
median&3.000&4.000\\%
\hline%
$\sigma$ (st. deviation)&2.651&2.700\\%
\hline\hline
\multicolumn{3}{c}{Wilcoxon test results}\\
\hline
p-value &\multicolumn{2}{c}{0.013}\\%
\hline%
\end{tabular}%
\caption{Results for the category: digital skills.}
\label{tab:category_dig_sk}
\end{table}

\subsection{Category: creativity}
\begin{figure}[ht!]
    \centering
    \includegraphics[width=\textwidth]{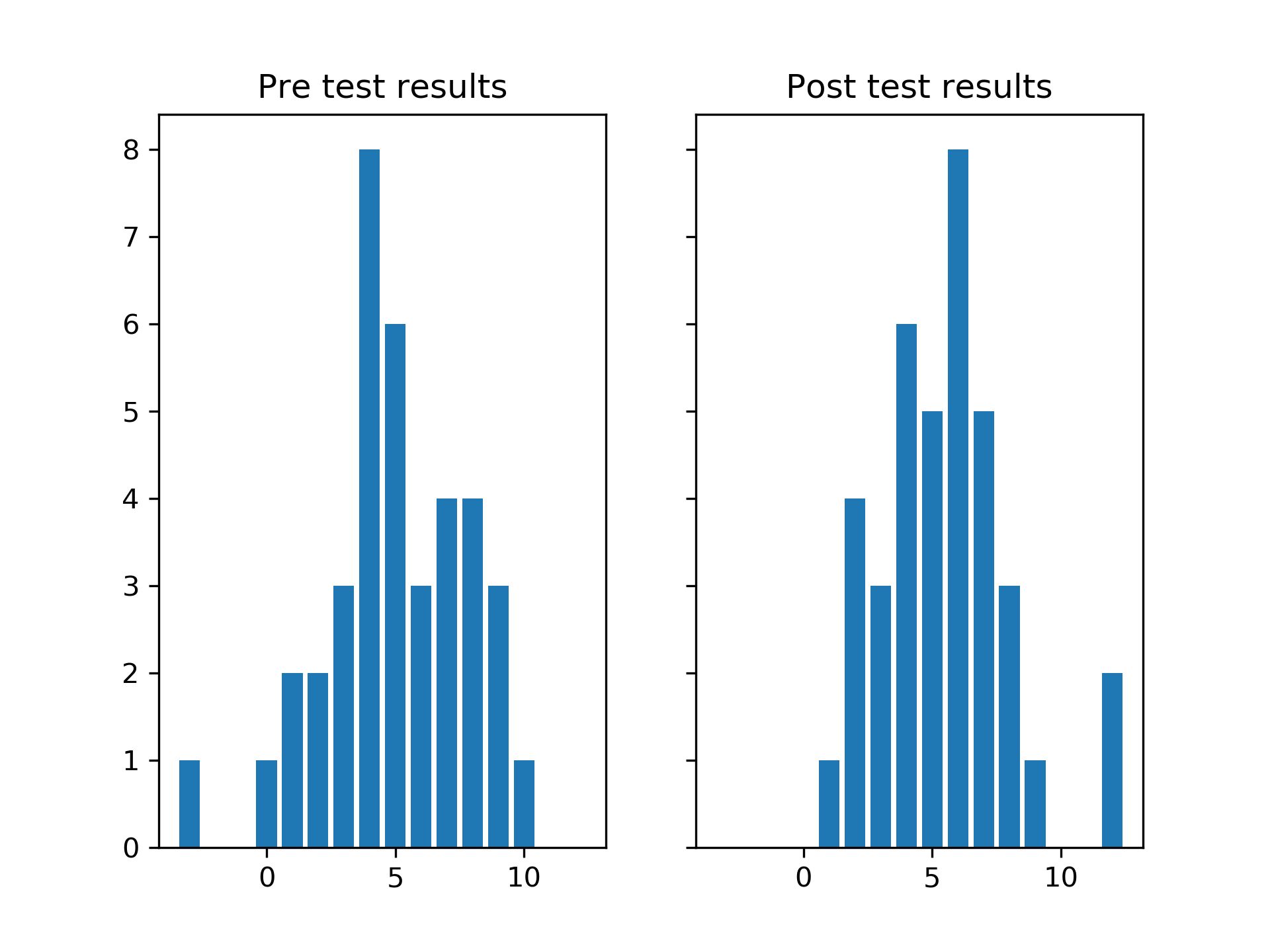}
    \label{fig:category_creat}
    \caption{Results for categories: creativity.}
\end{figure}

The category creativity contains six questions; we have proceeded the same way as in previous categories (see Fig.~\ref{fig:category_creat}, Tab.~\ref{tab:category_creat}).
% \begin{figure}[ht!]
%     \centering
%     \includegraphics[width=.4\textwidth]{cat_creat.jpg}
%     \caption{Results for the category: creativity.}
%     \label{fig:category_creat}
% \end{figure}

\begin{table}[ht!]
\centering
\begin{tabular}[c]{l|c|c}%
Parameter&pre&post\\%
\hline%
mean&4.974&5.447\\%
\hline%
median&5.000&5.500\\%
\hline%
$\sigma$ (st. deviation)&2.748&2.468\\%
\hline\hline
\multicolumn{3}{c}{Wilcoxon test results}\\
\hline
p-value &\multicolumn{2}{c}{0.095}\\%
\hline%
\end{tabular}%
\caption{Results for the category: creativity.}
\label{tab:category_creat}
\end{table}

The Wilcoxon test returned a p-value equal to $0.095$, so we cannot accept the $H_1$ hypothesis for this category.

\subsection{Category: critical thinking}
\begin{figure}[ht!]
    \includegraphics[width=\textwidth]{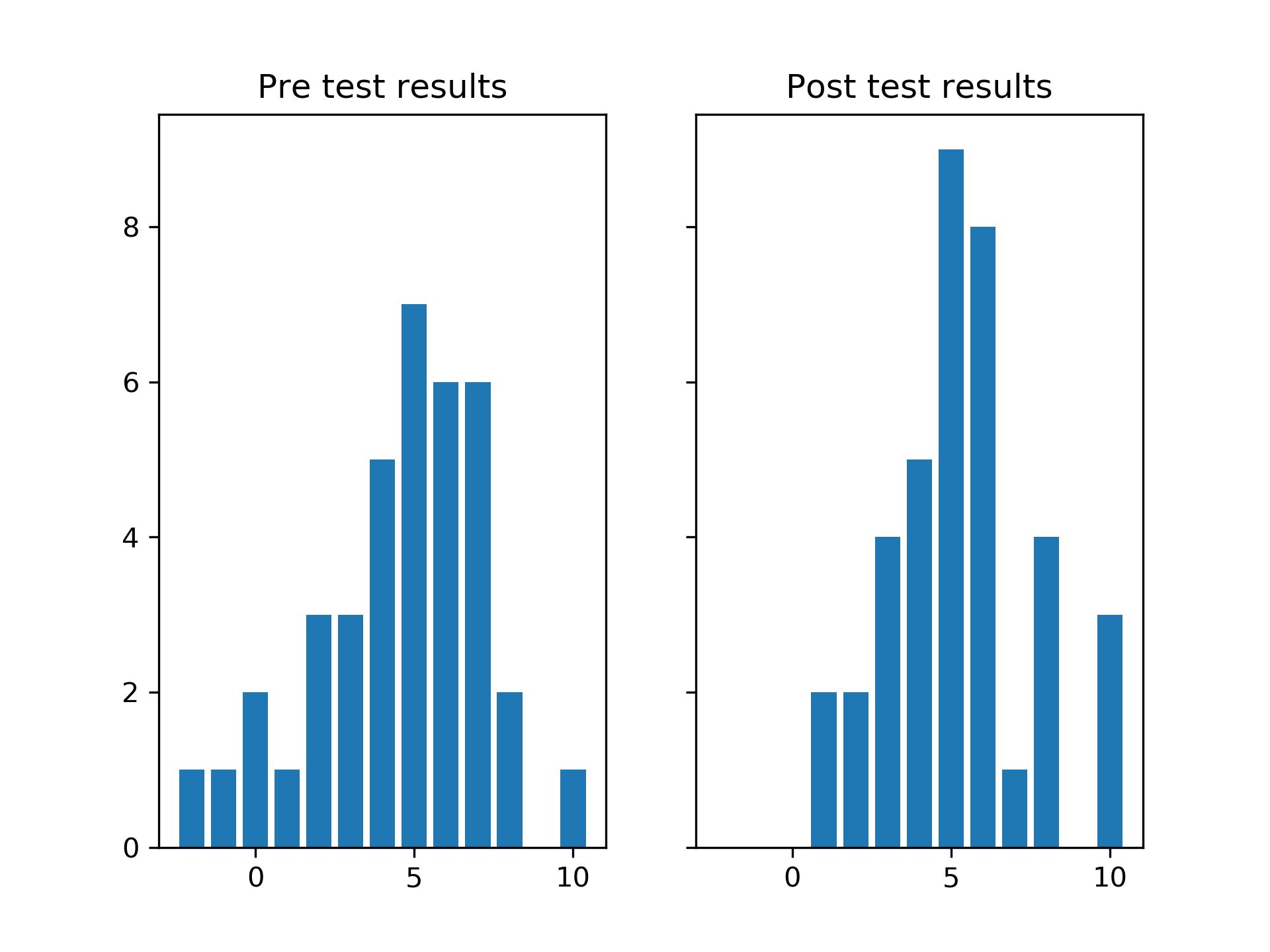}
    \label{fig:category_crit}
    \caption{Results for categories: creativity and critical thinking.}
\end{figure}

The category critical thinking contains five questions; we have proceeded in the same way as in previous categories (see Fig.~\ref{fig:category_crit} and Tab.~\ref{tab:category_crit}).
% \begin{figure}[ht!]
%     \centering
%     \includegraphics[width=.4\textwidth]{cat_crit.jpg}
%     \caption{Results for the category: critical thinking.}
%     \label{fig:category_crit}
% \end{figure}

\begin{table}[ht!]
\centering
\begin{tabular}[c]{l|c|c}%
Parameter&pre&post\\%
\hline%
mean&4.526&5.263\\%
\hline%
median&5.000&5.000\\%
\hline%
$\sigma$ (st. deviation)&2.613&2.244\\%
\hline\hline
\multicolumn{3}{c}{Wilcoxon test results}\\
\hline
p-value &\multicolumn{2}{c}{0.058}\\%
\hline%
\end{tabular}%
\caption{Results for the category: critical thinking.}
\label{tab:category_crit}
\end{table}

The Wilcoxon test returned a p-value equal to $0.058$, so we cannot accept the $H_1$ hypothesis for this category. The conclusion is that the distribution of pre and post-results are statistically similar.

\subsection{Category: willingness to learn}
\begin{figure}[ht!]
    \centering
    \includegraphics[width=\textwidth]{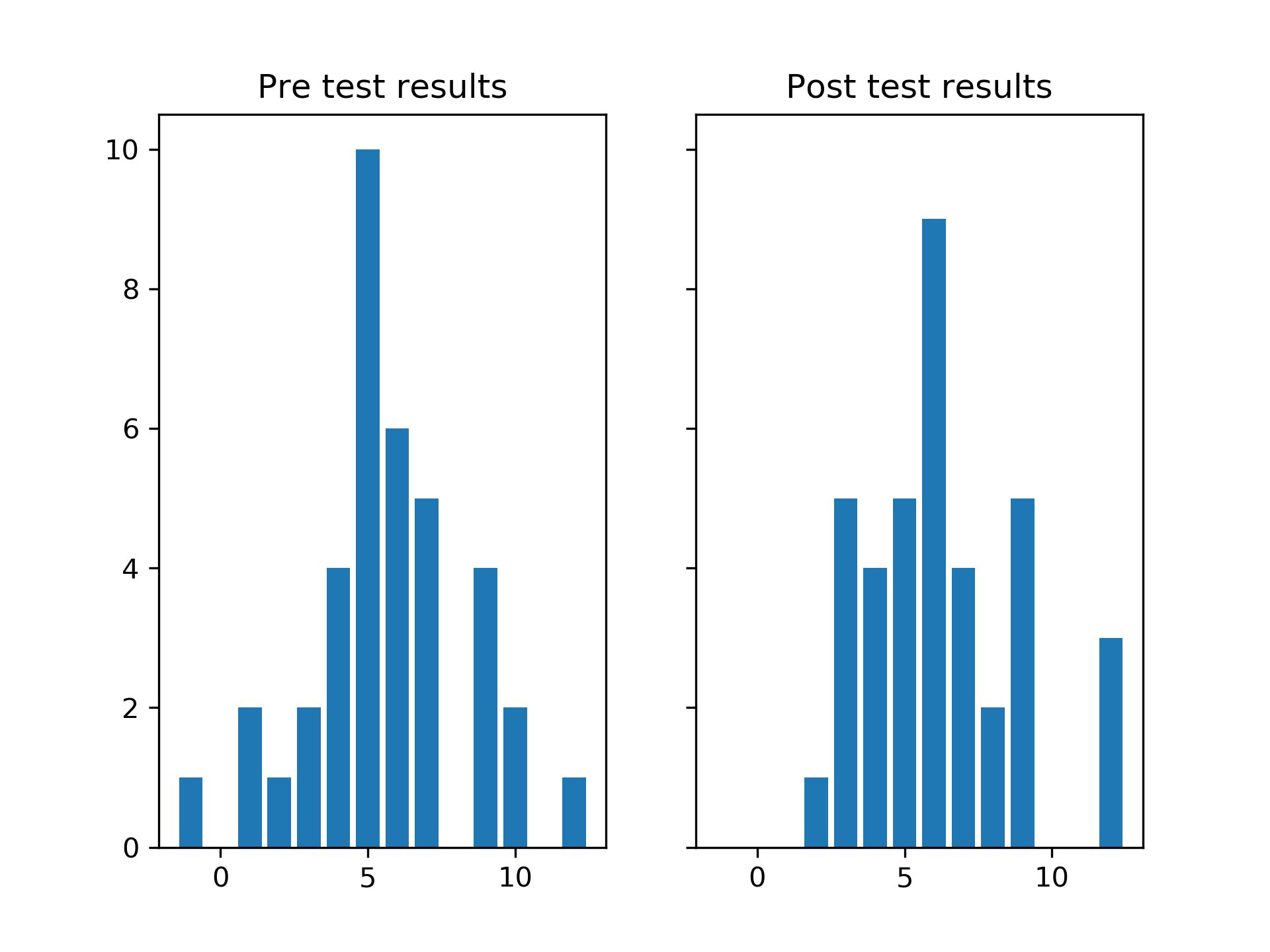}
    \label{fig:category_will}
    \caption{Results for categories: willingness to learn.}
\end{figure}

The category willingness to learn contains six questions (see Fig.~\ref{fig:category_will} and Tab.~\ref{tab:category_will}).
% \begin{figure}[ht!]
%     \centering
%     \includegraphics[width=.4\textwidth]{cat_will.jpg}
%     \caption{Results for the category: willingness to learn.}
%     \label{fig:category_will}
% \end{figure}

\begin{table}[ht!]
\centering
\begin{tabular}[c]{l|c|c}%
Parameter&pre&post\\%
\hline%
mean&5.632&6.237\\%
\hline%
median&5.000&6.000\\%
\hline%
$\sigma$ (st. deviation)&2.630&2.538\\%
\hline\hline
\multicolumn{3}{c}{Wilcoxon test results}\\
\hline
p-value &\multicolumn{2}{c}{0.073}\\%
\hline%
\end{tabular}%
\caption{Results for the category: willingness to learn.}
\label{tab:category_will}
\end{table}

The Wilcoxon test returned a p-value equal to $0.073$, so we cannot accept the $H_1$ hypothesis for this category. The conclusion is that the distribution of pre and post-results are statistically similar.

\subsection{Category: self reflection}
\begin{figure}[ht!]
\includegraphics[width=\textwidth]{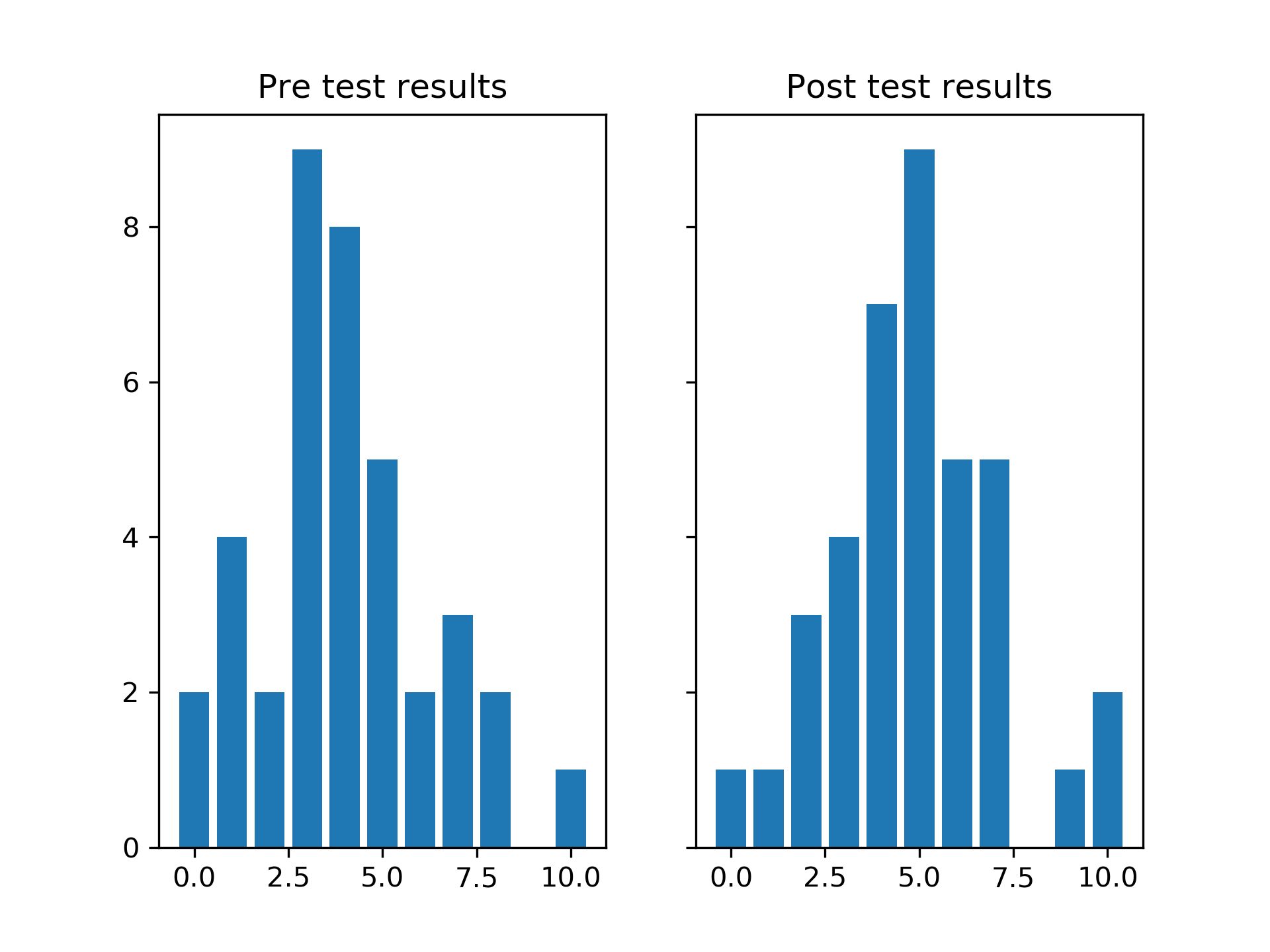}
    \label{fig:category_self_ref}
    \caption{Results for categories: flexibility and digital skills.}
\end{figure}

The category self reflection contains five questions (see Fig. \ref{fig:category_self_ref}, Tab.~\ref{tab:category_self_ref}).
% \begin{figure}[ht!]
%     \centering
%     \includegraphics[width=.4\textwidth]{cat_self_refl.jpg}
%     \caption{Results for the category: self reflection.}
%     \label{fig:category_self_ref}
% \end{figure}

\begin{table}[ht!]
\centering
\begin{tabular}[c]{l|c|c}%
Parameter&pre&post\\%
\hline%
mean&3.974&4.895\\%
\hline%
median&4.000&5.000\\%
\hline%
$\sigma$ (st. deviation)&2.253&2.186\\%
\hline\hline
\multicolumn{3}{c}{Wilcoxon test results}\\
\hline
p-value &\multicolumn{2}{c}{0.003}\\%
\hline%
\end{tabular}%
\caption{Results for the category: self reflection.}
\label{tab:category_self_ref}
\end{table}

The Wilcoxon test returned a p-value equal to $0.003$, and so we can accept the $H_1$ hypothesis for this category.

\section{Discussion}\label{sec:discussion}

% We can observe that for all categories the mean and median are higher in post- than in the pre-survey. However, it is not enough to look at these parameters as the samples are small and it is still possible that both came from the same statistics. The Wilcoxon statistical test gives more conservative answer.
The results obtained for each category are presented in Table \ref{tab:all_results}.
We can deduce that for four categories: communication, cooperation, digital skills, and self-reflection, the test results prove the increase of appropriate skills during the event. There is no reason to deduce similar implications for other categories like willingness to learn, critical thinking, flexibility, and creativity. In truth, the results are not very surprising. The event is only ten days long, and not all measured skills can be influenced in the same way during such a short time. The skills connected with communication, cooperation, and digital skills can be trained in a relatively short time. We think that self-reflection has increased as the participants discovered that they can obtain good results in a real-life task. The students learned that they can work and achieve required goals in a high-stress situation with emphasis on results. On the other hand, skills like creativity, flexibility, critical thinking, and willingness to learn can't be increased in short forms and require a longer process. It is worth to mention, that the applied tests do not decide if the $H_1$ hypothesis is improbable to be true. It shows that there is no reason to reject $H_0$. Additionally, all results are statistical, so in particular students the effects could be different.

\begin{table}[ht]
    \centering
    \begin{tabular}{c|c|c}
         Category   &  Wilcoxon& Conclusion\\
                    & test result&\\\hline
         Communication & $0.020$ & accept $H_1$ \\\hline
         Cooperation & $0.012$ & accept $H_1$\\\hline
         Flexibility & $0.117$ & reject $H_1$\\\hline
         Digital skills & $0.013$ & accept $H_1$\\\hline
         Creativity & $0.095$ &reject $H_1$\\\hline
         Critical thinking & $0.058$ &reject $H_1$\\\hline
         Willingness to learn & $0.073$ &reject $H_1$\\\hline
         Self reflection & $0.003$ & accept $H_1$\\\hline

    \end{tabular}
    \caption{Statistical results for all categories}
    \label{tab:all_results}
\end{table}

It is worth mentioning that there are differences between KYSS results and results obtained from direct questions from participants (see Tables \ref{tab:basic_questionaire} and \ref{tab:all_results}). For example, about $89\%$ of participants believe that their critical thinking was improved, but analysis of KYSS surveys did not prove that assessment. It could mean that the overall positive emotions to the event increase the percentage of positive answers. Additionally, the students are not always able to properly assess their skills.

\section{Conclusions}\label{sec:conclusions}

The method used in the project was expected to have a positive impact on the participants. Moreover, the event we describe is the second event of this type organized together. Staff members who have worked with students on the previous event observed positive impact on students' soft and future skills. The observations in the following academic year, suggest higher increase in skills in the project participants when compared with the other students on the degree programme. This time during the event, we have incorporated KYSS surveys as a measurement tool. The results support the expectations and previous observations. Not all aspects measured with KYSS show similar growth. The different effects of the project on different skill categories were expected. Moreover, the KYSS gives a more reliable assessment of the project's effects than simple direct questionnaire. Based on the presented results, the consortium will conduct a similar effect and measure its impacts on the participants.

The effects, show that Software Develompment projects can be used for multidisciplinary education. This type of team task allows to increase communication and cooperation in diverse group. Young people are eager to use and create new solutions that use modern technologies. What is important that this type of activities is very appreciated by non-technical students.

The results prove that the intensive international and multidisciplinary projects can lead to significant impact on participants soft and future skills. Moreover, our research provide for the first time objective measurements of the effects of the MIMI methodology.

\section*{Acknowledgment}
This work has been (partially) funded by the ERASMUS+ program of the European Union under grant no. 2021-1-DE01-KA220-HED-000023215.
Neither the European Commission nor the project’s national funding agency DAAD are
responsible for the content or liable for any losses or damage resulting of the use of
these resources.

\bibliographystyle{plain}
\bibliography{manuscript}


\end{document}